# Determination of the Electric Dipole Moment of a Molecule from Density Functional Theory Calculations

**Byeong June Min**

***Department of Physics, Daegu University, Kyungsan 712-714, Korea***

Density functional theory (DFT) calculation has had huge success as a tool capable of predicting important physical and chemical properties of condensed matter systems. We calculate the electric dipole moment of a molecule by using the differential electron density with respect to the superimposed electron density of the free atoms, exploiting the cancellation of DFT errors. Our results on a range of molecules show an excellent agreement with experiments.

Email: bjmin@daegu.ac.kr

Fax: +82-53-850-6439, Tel: +82-53-850-6436

Density functional theory (DFT) calculations can accurately calculate many important physical and chemical properties of a condensed matter system. However, the DFT results of the electric dipole moment of a molecule show an unusually wide margin of error, sometimes approaching a factor of 2 depending on the choice of the basis functions used in the calculation [1]. Such a situation is puzzling in view of the DFT's consistent success on other fronts. Furthermore, DFT calculations produce the electron density of the material with good reliability and, thus, everything needed to determine the electric dipole moment with accuracy.

The most obvious way to remove this dependency on the choice of the basis functions would be to use plane-wave basis. However, study of Nb clusters by K. E. Andersen *et al*. [2] was the only plane-wave DFT study on the electric dipole moment we could find. The authors used the self-consistent output electron density $\rho(\vec{r})$ to calculate the dipole moment as $\vec{P} = -e\int \vec{r}\rho(\vec{r}) d^3\vec{r} + e\sum Z_i \vec{R}_i$ . Instead, we used the differential electron density $\rho_{diff}(\vec{r}) = \rho(\vec{r}) - \sum \rho_{atom}(\vec{r})$ to calculate the dipole moment as $\vec{P} = -e\int \vec{r}\rho_{diff}(\vec{r}) d^3\vec{r}$ , exploiting the cancellation of the DFT errors.

We have used ABINIT package [3] for the present calculation with periodic boundary condition and gamma point sampling. Norm-conserving pseudopotentials [4] were used. Plane wave energy cutoff of 1200 eV is used. The cubic box dimension was chosen as 13.2 Å. The total-energy converges within 1 meV. Exchange correlation energy was described by the Perdew-Burke-Ernzerhof parameterization within the generalized gradient approximation (PBE-GGA) [5]. Semi-empirical treatment of the van der Waals interaction is used [6].

Self-consistency cycles were repeated until the difference of the total energy becomes smaller than $2.7 \times 10^{-6}$ eV, twice in a row. The system was relaxed until the average force on the atoms becomes smaller than $2.3 \times 10^{-3}$ eV/A by Broyden-Fletcher-Goldfarb-Shanno (BFGS) minimization scheme [7].

We selected to work on several diatomic molecules and a few small molecules listed in the NIST Computational Chemistry Comparison and Benchmark Database [1]. For example, we obtained 3.97 Debye for CaCl that is often used as a test case of alkaline-earth monohalides. W. E. Ernst *et al*. [8] reported a dipole moment of 4.257 Debye by molecular beam laser-microwave double-resonance method and compared the result with 3.6 Debye from theoretical studies [9-12]. In the case of formaldehyde oxime ($CH_2NOH$), straightforward calculation yielded 0.28 Debye against 0.44 Debye from experiment. For its isomer, formamide ($HCONH_2$), we obtained 3.91 Debye against 3.73 Debye from experiment.

As a comparison, using the Hirshfeld charges $q_H$ of the atoms in the molecule, the electric dipole moments are calculated by $\vec{P} = \sum q_{H,i}\vec{R}_i$ . This approach yields qualitative agreement with the experiment. We obtain 4.32 Debye for CaCl, 0.13 Debye for $CH_2NOH$, 2.8 Debye for $HCONH_2$, and 5.67 Debye for NaF. It discerns the isomers $CH_2NOH$ and $HCONH_2$ to some degree. The reason for the qualitative agreement may be found from the fact that Hirshfeld charges are determined by calculating the deviation of the charge distribution from the charge distribution of the free atom.

The results are summarized in Table 1 and Figure 1, together with the results from the Hirshfeld charge method and experiment. Results for CaF, another alkaline-earth monohalide, is added at the end of Table 1 along with experimental result [13]. The agreement with experiment is typically within a few percent, with occasional exceptions.

The differential electron density $\rho(\vec{r}) - \sum \rho_{atom}(\vec{r})$ now may be viewed as the dipole charge distribution because the stationary atom can only be charge neutral. The differential electron density isosurface for CaCl is shown in Figure 2.

Because the key point of the present method is taking the differential charge density from the superimposed atomic charge density, it can be readily applied to calculations that use local orbital basis. It would be interesting to see how much difference will be made by this method.

To summarize, we propose an effective method for calculating the electric dipole moment of a molecule by using the differential charge density instead of the direct output density from a self-consistent DFT calculation.

This research was supported in part by the Daegu University Research Funds.

Table 1. Electric dipole moment (Debye) of selected molecules calculated by using Hirshfeld charges, and by using the differential charge density.

| | Experiment | Hirshfeld | Differential Charge Density |
|---|---|---|---|
| CaCl | 4.26 | 4.32 | 3.97 |
| LiH | 5.88 | 3.19 | 5.71 |
| BH | 1.27 | 0.39 | 1.52 |
| CN | 1.45 | 0.6 | 1.18 |
| OH | 1.66 | 0.71 | 1.61 |
| LiO | 6.84 | 3.88 | 6.11 |
| CO | 0.11 | 0.34 | 0.22 |
| NO | 0.153 | 0.005 | 0.24 |
| HF | 1.82 | 0.91 | 1.78 |
| LiF | 6.28 | 4.28 | 6.16 |
| NaF | 8.12 | 5.67 | 7.88 |
| AlF | 1.53 | 2.24 | 1.57 |
| SiO | 3.1 | 2.45 | 3.03 |
| PN | 2.75 | 1.62 | 2.79 |
| PO | 1.88 | 1.59 | 1.87 |
| SiS | 1.73 | 1.77 | 1.65 |
| CS | 1.98 | 0.67 | 2.02 |
| NS | 1.81 | 1.04 | 1.82 |
| SO | 1.55 | 1.14 | 1.44 |
| $NH_2OH$ | 0.59 | 0.45 | 0.57 |
| HCOOH | 1.41 | 1.26 | 1.5 |
| $HNO_3$ | 2.17 | 1.25 | 2.07 |
| $CH_3OH$ | 1.7 | 0.78 | 1.61 |
| $CH_2NOH$ | 0.44 | 0.13 | 0.28 |
| $HCONH_2$ | 3.73 | 2.8 | 3.91 |
| CaF | 3.07 | 3.77 | 2.88 |

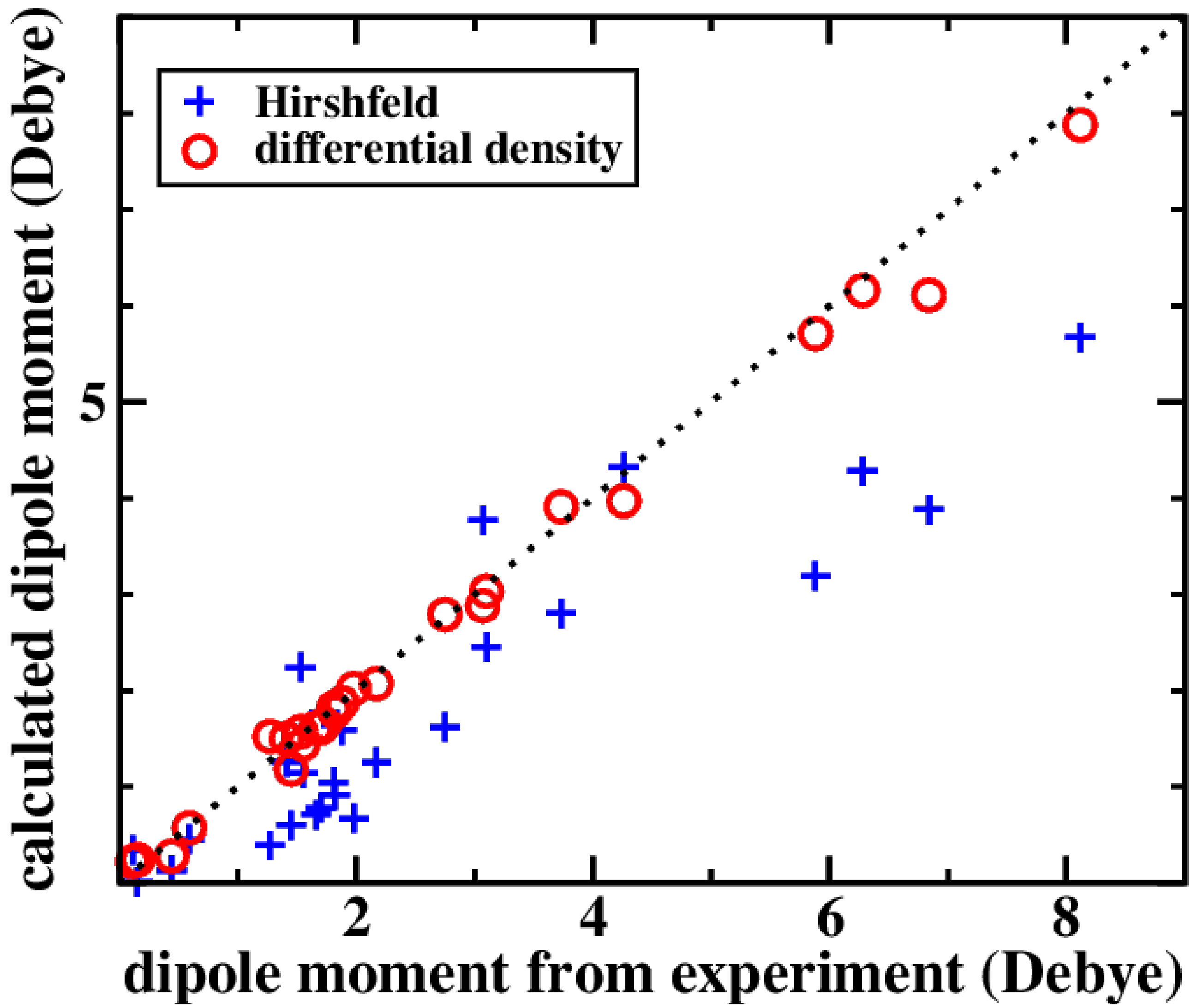

Fig. 1.

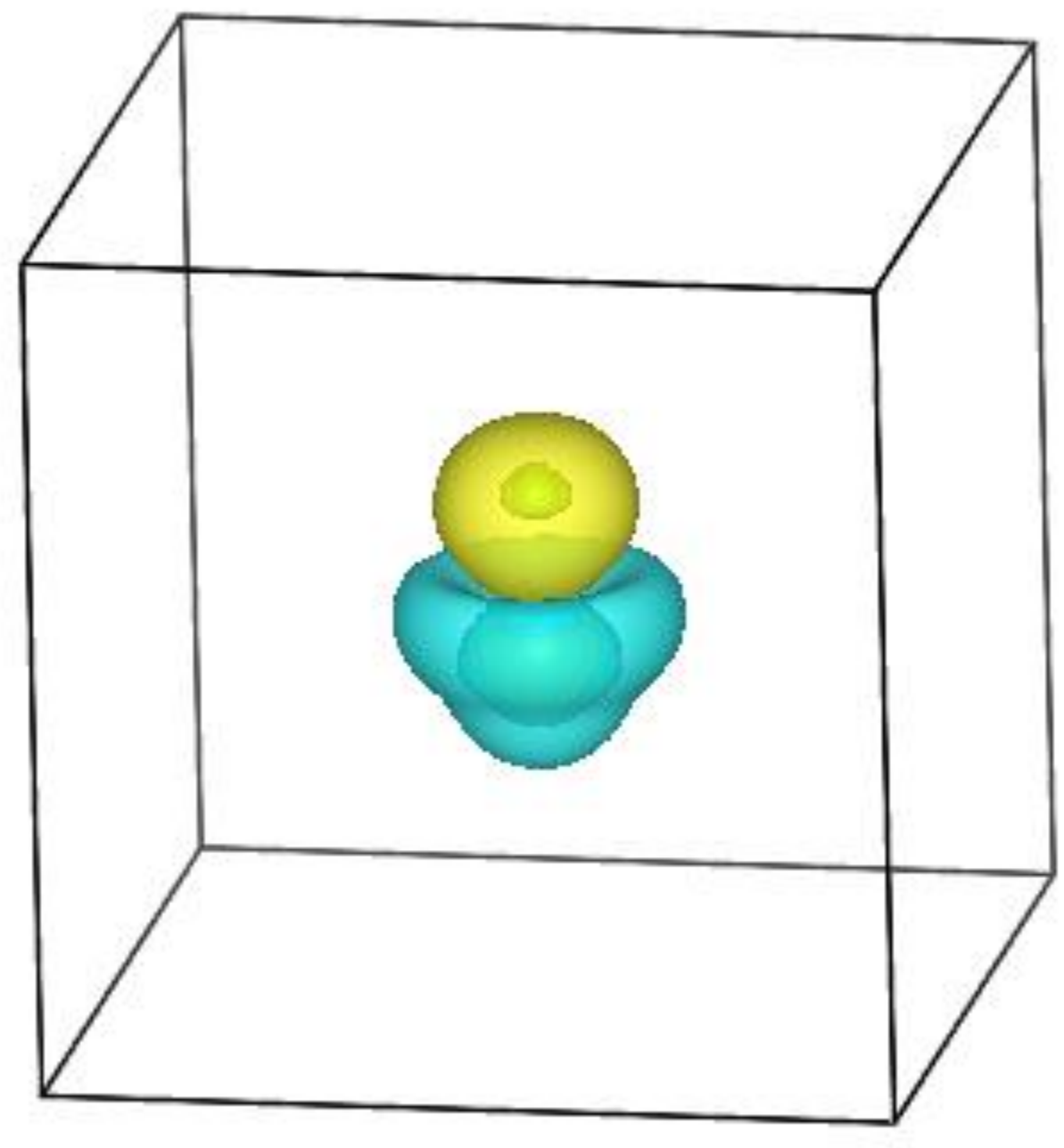

Fig. 2

Figure Captions.

Fig. 1. Calculated versus experimental electric dipole moment of molecules in Table 1. Hirshfeld charge method has a tendency to underestimate the electric dipole moment. When the differential electron density with respect to the neutral atom is used, the agreement between the theory and experiment becomes excellent.

Fig. 2. The differential electron density isosurface for CaCl. This may be viewed as the dipole charge density distribution because the stationary atom is charge neutral. Ca is surrounded by the blue isosurface, meaning decrease of the electron density.